\documentstyle[12pt,twoside,fleqn,espcrc1,epsfig,glas]{article}
\newlength{\dinwidth}
\newlength{\dinmargin}
\def\lsim{\mathrel{\rlap{\lower4pt\hbox{\hskip1pt$\sim$}}
    \raise1pt\hbox{$<$}}}                
\def\gsim{\mathrel{\rlap{\lower4pt\hbox{\hskip1pt$\sim$}}
    \raise1pt\hbox{$>$}}}                
\begin{document}

\begin{titlepage}{GLAS-PPE/1999-25}{16$^{\underline{\rm{th}}}$ December 1999}
\title{NLO Program Comparison for Event Shapes}

\author{G.J. McCance
}


\vspace{0.3cm}
\centerline{\em Work presented at the 1999 HERA Monte Carlo Workshop}

\vspace{0.8cm}
\begin{abstract}
The analysis of event shape variables enable studies of QCD and determinations of $\alpha_s$, but require predictions to next-to-leading order. The predictions of two next-to-leading order programs are compared for current jet thrust, current jet-broadening, $C$ parameter and jet-mass in deep inelastic scattering.
\end{abstract}
\end{titlepage}

\section{Introduction}

The large number of deep inelastic scattering (DIS) events collected at HERA permit the study of hadronic final states to high precision. Event shape variables describe the shape of jets of hadrons that are produced by the fragmenting quark. These have been investigated in $e^{+}e^{-}$ experiments, and can be used to extract the strong coupling constant independent of a particular jet-finding algorithm. Interest has recently been revived by the calculation of power corrections\cite{paper:powerevent}. A similar analysis can be performed in DIS and compared to the $e^{+}e^{-}$ data.

A number of general NLO programs have recently become available. The work presented here compares two programs, DISENT\cite{manual:disent} and DISASTER++\cite{manual:disaster}, interfaced through a common control library\cite{manual:commonlib}. Both programs use the subtraction method\cite{paper:bigsubtract} and allow arbitrary (2+1) jet observables to be evaluated in the Breit frame of reference. The two programs were compared for both differential and mean distributions over a wide range in $x$ and $Q^2$ to match with an existing ZEUS analysis\cite{thesis:rob}.

\section{Event Shape Variable Definition}

The event shape variables considered were current jet thrust $T_z$\footnote{It should be noted that this is different from the standard definition of thrust $T$ where the thrust axis has to be {\em searched} for by maximising the thrust in the expression. For $T_z$, the thrust axis is defined by the virtual boson axis.}, current jet broadening $B_c$, invariant jet mass $\rho_E$ and the $C$ parameter. The variables are defined so as to be infrared and collinear safe to avoid divergences in the Monte Carlo integration and are calculated at parton level.

\begin{equation}
T_z = \frac{\sum_{i}{|\vec{n}.\vec{p}_{i}|}}{\sum_{i}{|\vec{p}_{i}|}}
    = \frac{\sum_{i}{p_z}}{\sum_{i}{|\vec{p}_{i}|}}    \label{eqn:tz}
\end{equation}

\begin{equation}
B_c = \frac{\sum_{i}{|\vec{n}\times\vec{p}_{i}|}} {2\sum_{i}{|\vec{p}_{i}|}}
    = \frac{\sum_{i}{p_{\bot}}}{2\sum_{i}{|\vec{p}_{i}|}}   \label{eqn:jbroad}
\end{equation}

\begin{equation}
\rho_E = \frac{M^{2}}{{(2 E_{tot})}^{2}}
       = \frac{ {(\sum_{i} p^{\mu}_{i})}^{2} } { 4{(\sum_{i} E_{i})}^{2} }    \label{eqn:jmass}
\end{equation}

\begin{equation}
C = 3(\lambda_{1}\lambda_{2} + \lambda_{2}\lambda_{3} + \lambda_{1}\lambda_{3})
 \label{eqn:cpar}                                     
\end{equation}

where $E_{tot}$ is the total energy of all the particles in the current region. $\lambda_{1,2,3}$ are the eigenvalues of the linearised momentum tensor, $\Theta^{\alpha,\beta}$, whose components are given by:

\begin{equation}
\Theta^{\alpha,\beta} = \frac{ \sum_{i} \vec{p}^{\alpha}_{i} \vec{p}^{\beta}_{i} / |\vec{p}_{i}| }
                             { \sum_{i} |\vec{p}_{i}| }
\end{equation}

The sum $i$ is over all the partons in the current region of the Breit frame. The particles' three-momenta are given by $\vec{p}_{i}$, and their energies by $E_{i}$. The momentum four-vector of the particles is denoted $p^{\mu}_{i}$, and $\vec{n}$ is a unit vector aligned along the virtual boson axis ($n = [0,0,-1]$).

In the Born approximation corresponding to the quark parton model (QPM), all the above variables are equal to zero, except thrust which is equal to unity. For consistency with other variables, it is convenient to define $\tau_z = 1 - T_z$, which goes to zero in the Born approximation.

\section{Program Comparison}
For both programs, DISENT and DISASTER++, events were generated in 17 kinematic bins of $x$ and $Q^2$ as shown in Table \ref{tabl:kinema}.
\begin{table}[htb]
\caption{\it Kinematic bins used in the comparison.}
\begin{center}
\begin{tabular}{|c|c|c||c|c|c|}
\hline
Bin & $x_{min} - x_{max}$ & $Q^{2}_{min} - Q^{2}_{max}$ & Bin & $x_{min} - x_{max}$ & $Q^{2}
_{min} - Q^{2}_{max}$\\
\hline \hline
1 & 0.0006 - 0.0012 & 10.0 - 20.0 & 9 & 0.0024 - 0.0100 & 160.0 - 320.0\\
2 & 0.0012 - 0.0024 & 10.0 - 20.0 & 10 & 0.0100 - 0.0500 & 160.0 - 320.0\\
3 & 0.0012 - 0.0024 & 20.0 - 40.0 & 11 & 0.0100 - 0.0500 & 320.0 - 640.0\\
4 & 0.0024 - 0.0100 & 20.0 - 40.0 & 12 & 0.0100 - 0.0500 & 640.0 - 1280.0\\
5 & 0.0012 - 0.0024 & 40.0 - 80.0 & 13 & 0.0250 - 0.1500 & 1280.0 - 2560.0\\
6 & 0.0024 - 0.0100 & 40.0 - 80.0 & 14 & 0.0500 - 0.2500 & 2560.0 - 5120.0\\
7 & 0.0024 - 0.0100 & 80.0 - 160.0 & 15 & 0.0600 - 0.4000 & 5120.0 - 10240.0\\
8 & 0.0100 - 0.0500 & 80.0 - 160.0  & 16 & 0.1000 - 0.6000 & 10240.0 - 20480.0\\
&  &              & 17 & 0.2000 - 0.8000 & 20480.0 - 40960.0\\
\hline
\end{tabular}
\label{tabl:kinema}
\end{center}
\end{table}                   
The matrix elements produced by the programs are convoluted with a parton density function (PDF) and the value of $\alpha_s$ evaluated at the relevant scale. The CTEQ-4M \cite{paper:cteq4m} parton density function was used based on the  $\overline{MS}$ renormalisation scheme. The renormalisation and factorisation scales, $\mu_R$ and $\mu_F$, were set to the momentum transfer $Q$.

To ensure that the shape variables are infrared and collinear safe, an energy cut must be imposed so that the Monte Carlo integration does not diverge\cite{private:3percent}. It is required that at least $3\%$ of the total energy available to the current region, $Q/2$, has actually gone into the current region. 

\begin{figure}[p]
\centering
\scalebox{0.90}{\epsfig{file=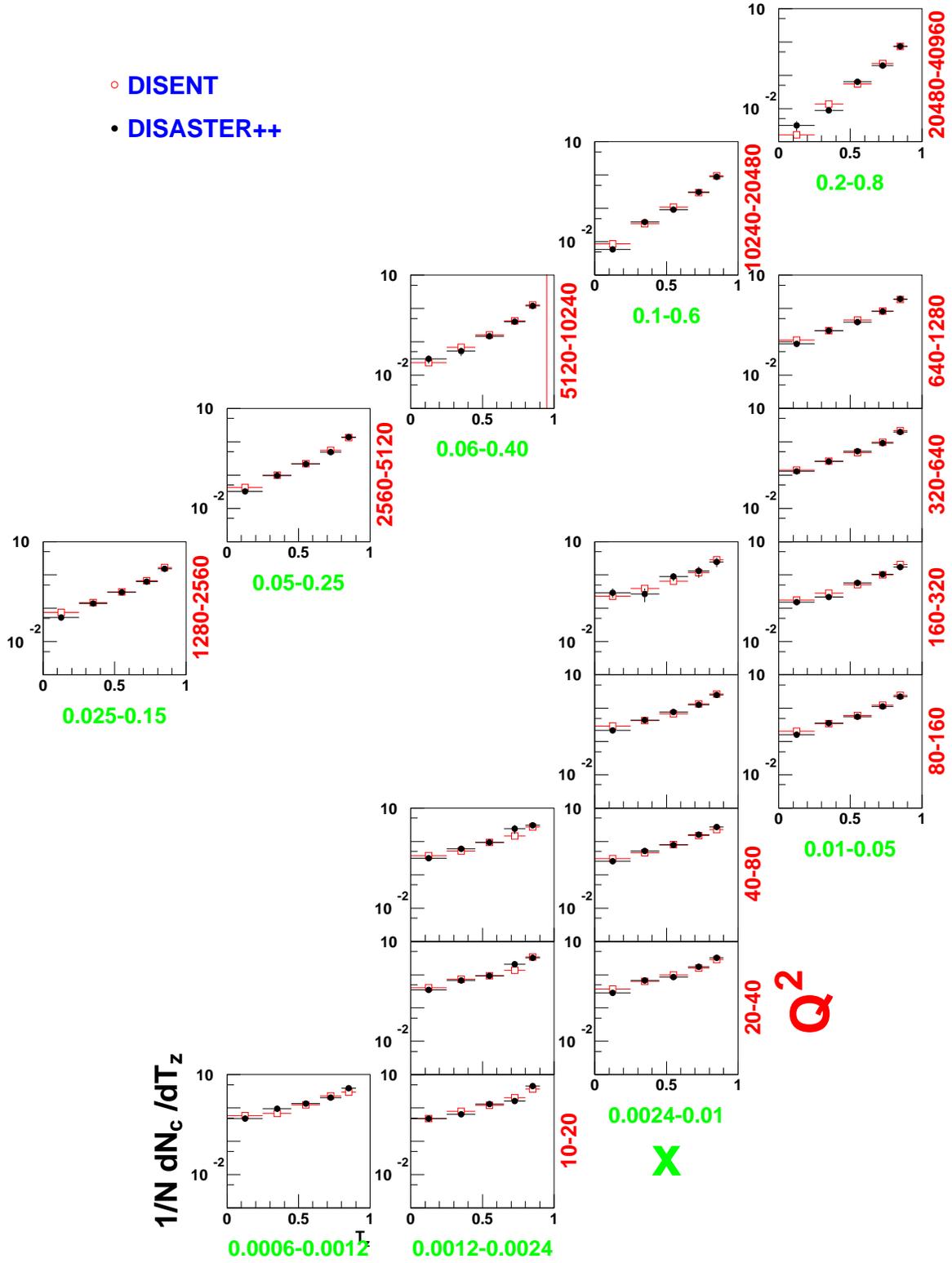,width=\the\textwidth}}
\caption{\it $T_z$ differential distribution.}
\label{fig:difftz}
\end{figure}

\begin{figure}[p]
\centering
\scalebox{0.90}{\epsfig{file=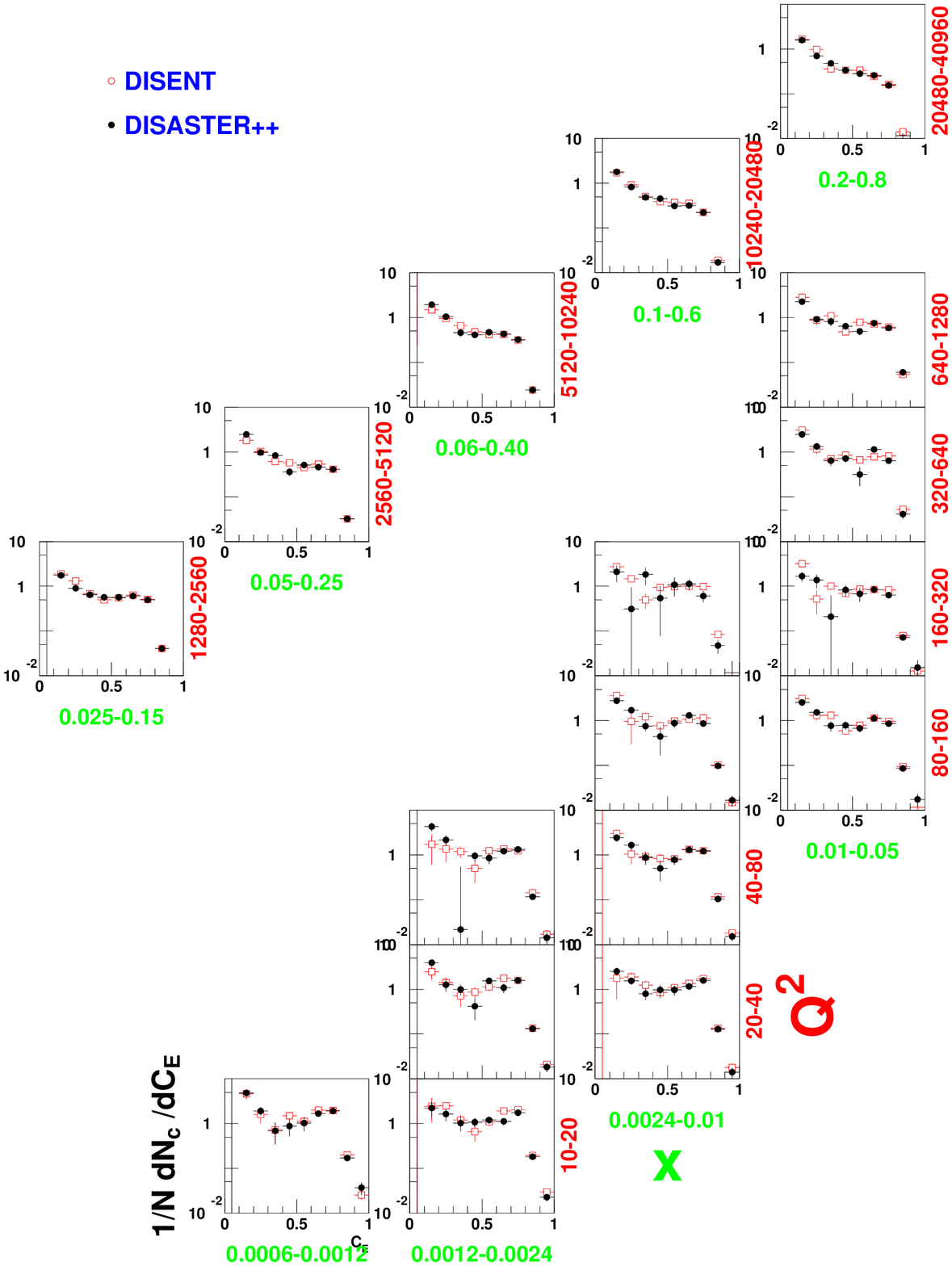,width=\the\textwidth}}
\caption{\it $C$ parameter differential distribution.}
\label{fig:diffcpar}
\end{figure}

\begin{figure}[p]
\centering
\scalebox{0.90}{\epsfig{file=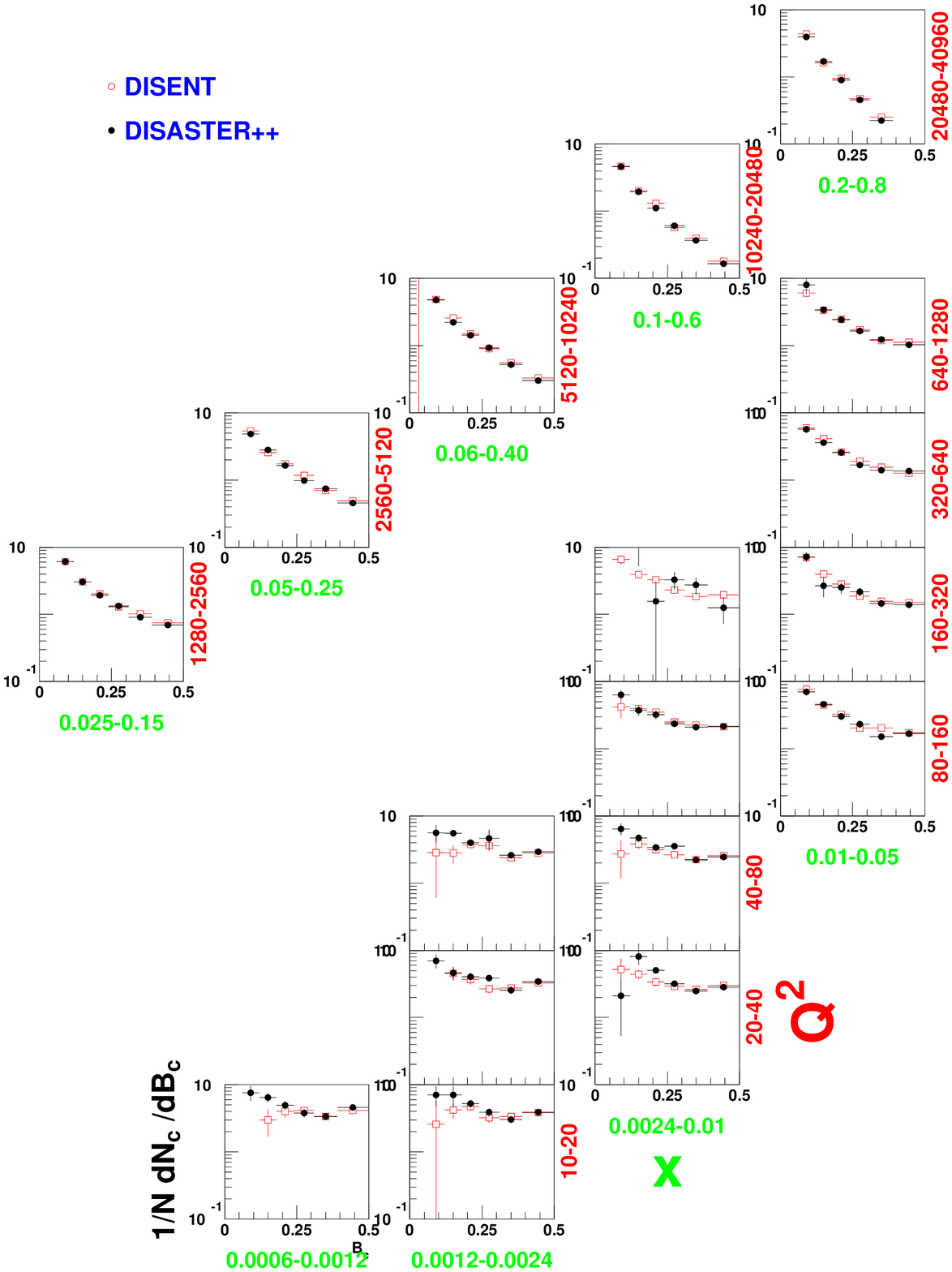,width=\the\textwidth}}
\caption{\it $B_c$ differential distribution.}
\label{fig:diffjbroad}
\end{figure}

\begin{figure}[p]
\centering
\scalebox{0.90}{\epsfig{file=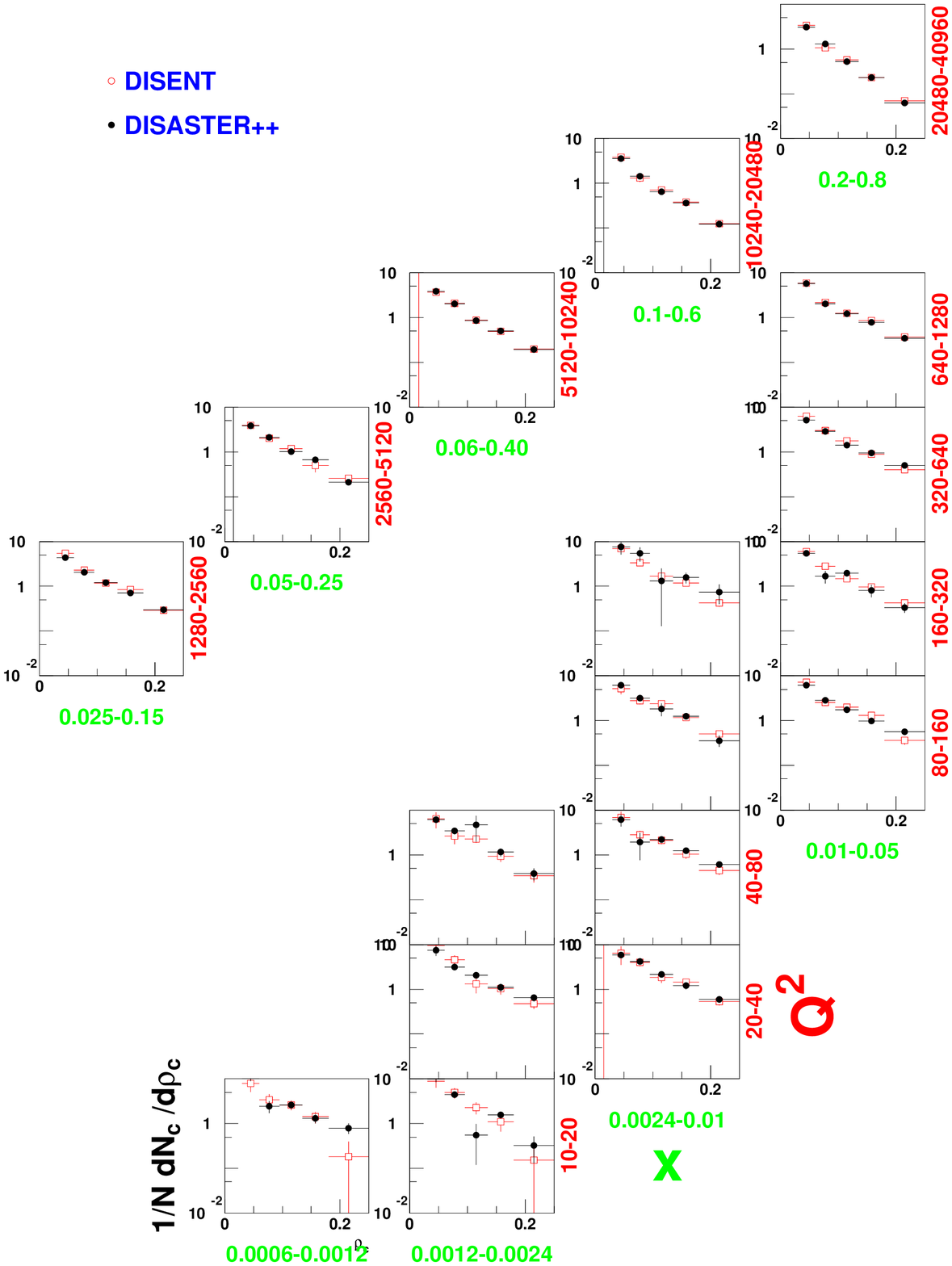,width=\the\textwidth}}
\caption{\it $\rho_c$ differential distribution.}
\label{fig:diffjmass}
\end{figure}

Figures \ref{fig:difftz} to \ref{fig:diffjmass} show the differential distributions for the four variables. The kinematic bins are arranged so that $x$ increases going from left to right, and $Q^2$ increases going from bottom to top. The bin widths were chosen to match an existing ZEUS analysis.

\begin{itemize}
\item For thrust in figure \ref{fig:difftz}, a clear evolution in $Q^2$ is visible where as $Q^2$ increases, the high thrust population increases as the events become more collimated. Both programs agree very well over the whole kinematic range, and are consistent within statistical errors. In the high thrust region, where the events become very collimated, both programs encounter collinear singularities. In this region the calculations diverge (the highest thrust bin $0.9 \leq T_z \leq 1.0$ has a negative value indicating divergence).
\item The $C$ parameter distributions in figure \ref{fig:diffcpar} show an increasing tendency to favour (2+1) jet-like events with increasing $Q^2$. The expected fall off\cite{paper:resummedc} at $C=\frac{3}{4}$ is reproduced by both programs. There is reasonable agreement between the two programs, particularly in the high $Q^2$ region. The same collinear singularities that cause thrust to diverge are visible as $C\rightarrow 0$; this appears either as a divergent point or a point with a large error. For DISASTER++ a few kinematic bins show a point in the middle of the distribution with a large error indicative of some divergent behaviour; this is currently being investigated by the author of DISASTER++.
\item Jet-broadening in figure \ref{fig:diffjbroad} shows the expected behaviour; the broadness of the jet reduces as $Q^2$ increases. The programs agree well in the high $Q^2$ region, however below a $Q^2$ of 100 GeV$^2$, there is some disagreement.
\item Figure \ref{fig:diffjmass} shows good agreement for the jet-mass in most of the kinematic bins, although some of the medium $Q^2$ bins ($160 - 320$ GeV$^2$) by around 10\% level. The errors in the low $Q^2$ regions are comparatively large indicating some problem in the numerical stability of the Monte Carlo integration in these regions for this observable. The numerical stability can, in general, be improved by increasing the number of points generated in each kinematic bin. 
\end{itemize}

\begin{figure}[htb]
\centering
\scalebox{.45}{\epsfig{file=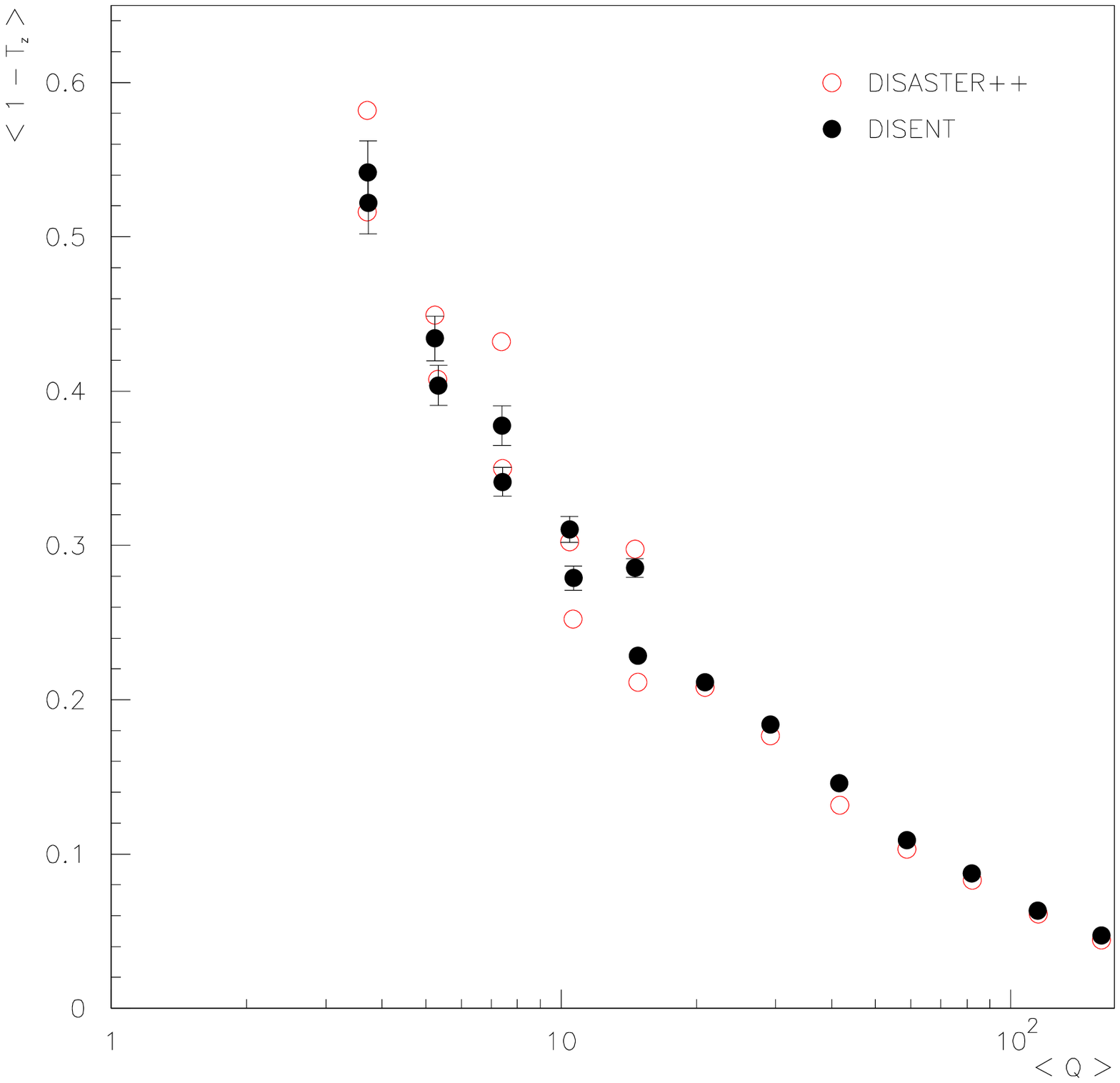,width=\the\textwidth}}
\scalebox{.45}{\epsfig{file=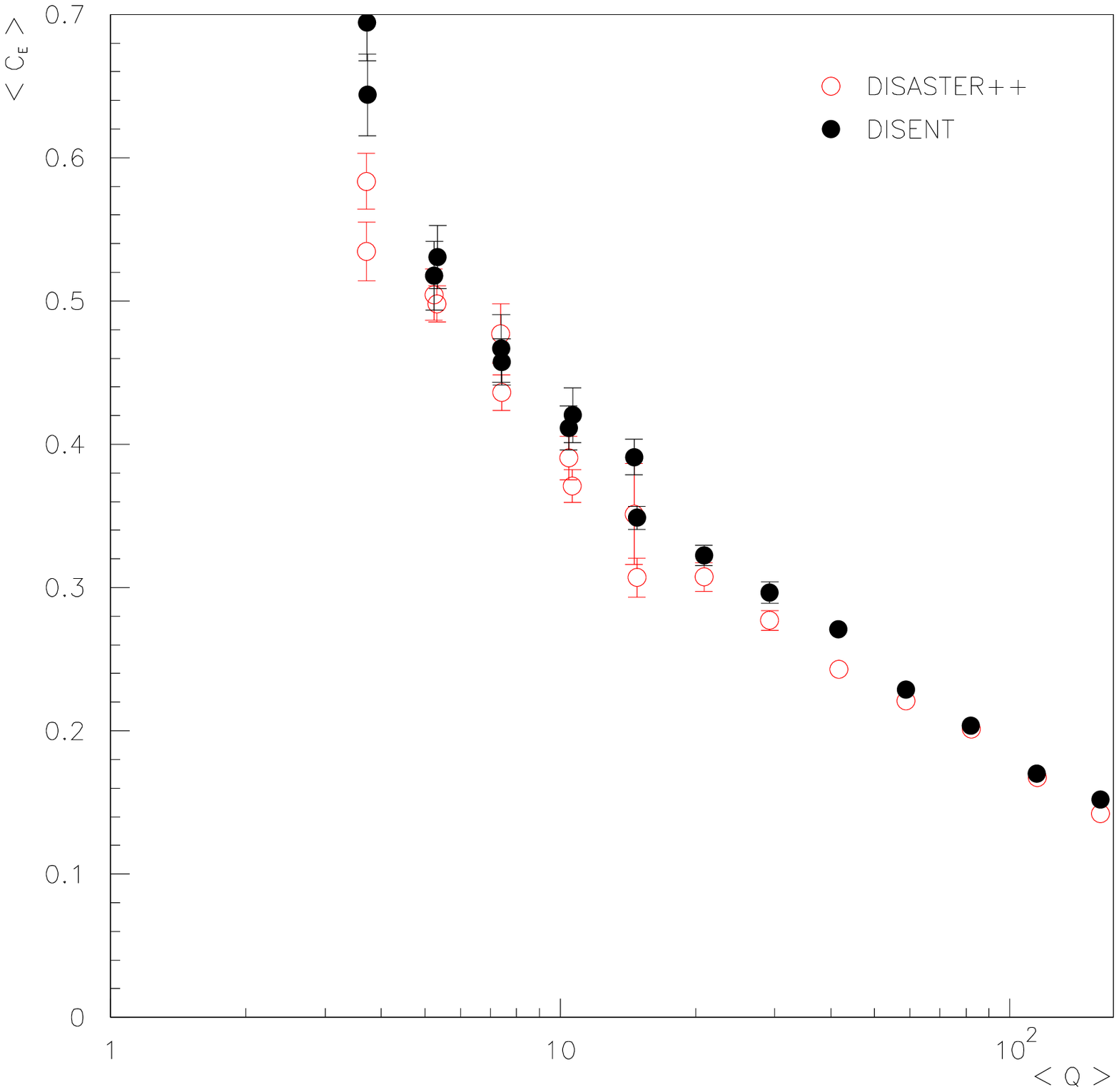,width=\the\textwidth}}
\scalebox{.45}{\epsfig{file=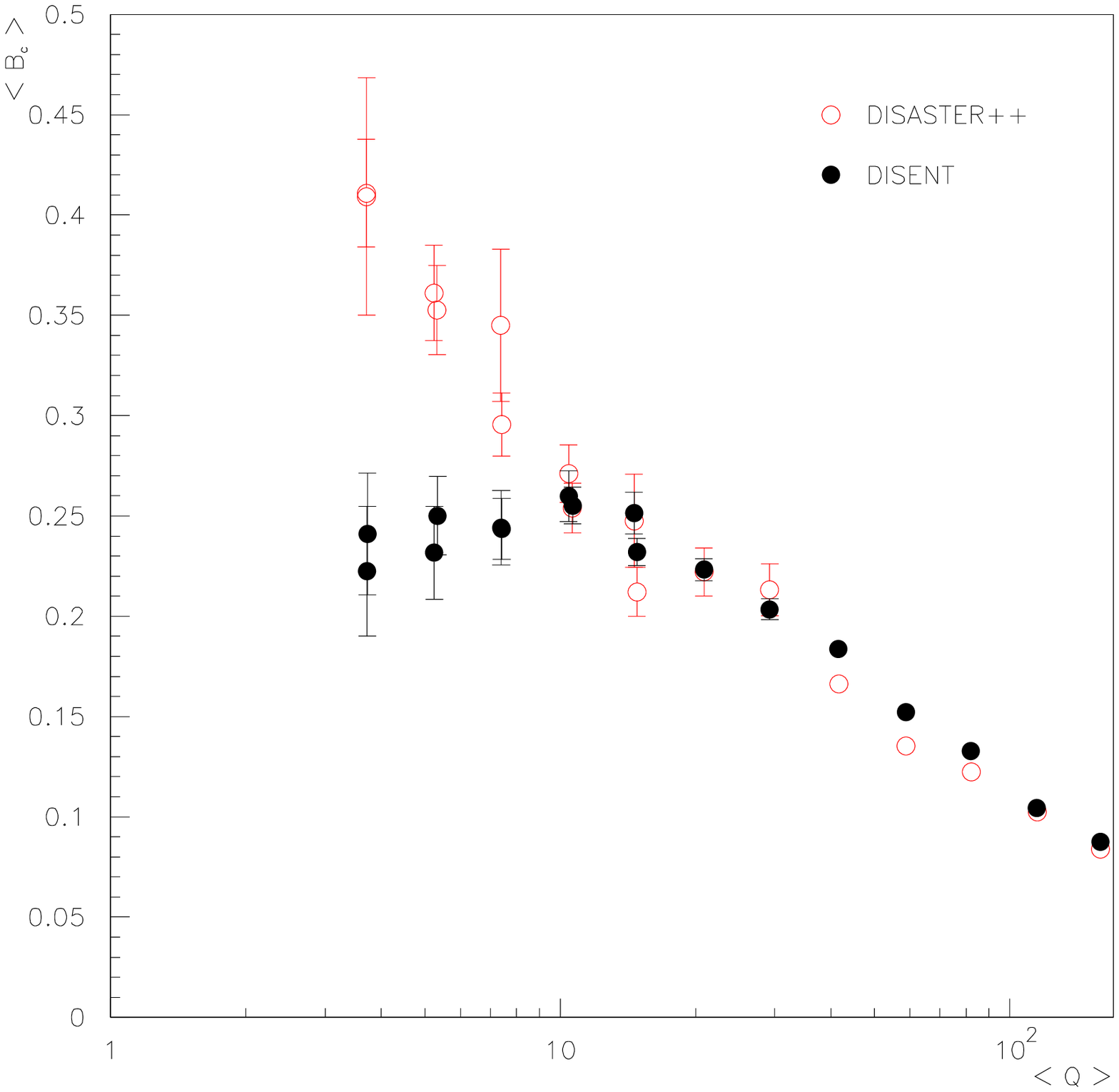,width=\the\textwidth}}
\scalebox{.45}{\epsfig{file=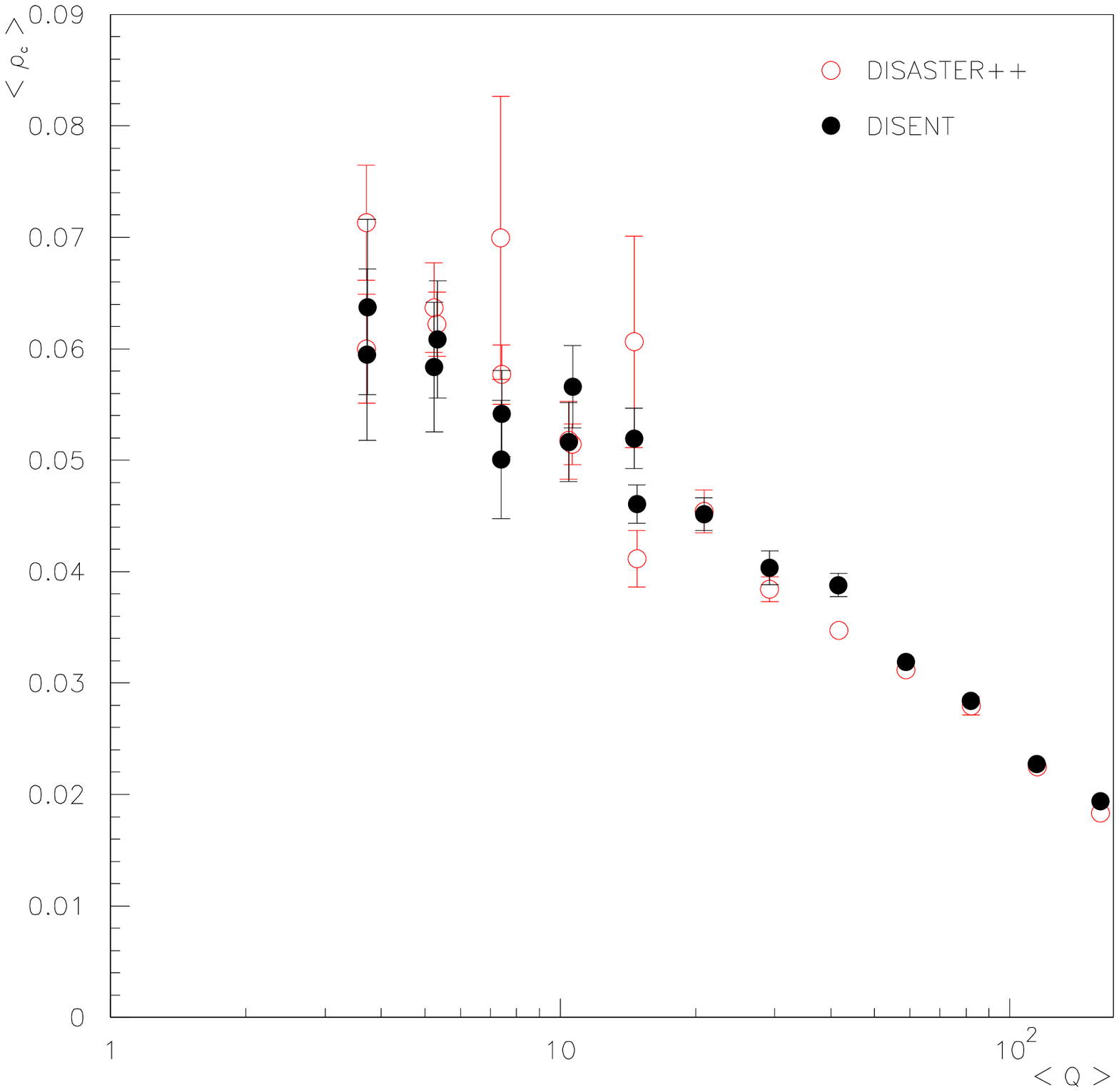,width=\the\textwidth}}
\caption{\it Mean event shape distributions vs. $<Q>$.}
\label{fig:means}
\end{figure}

Figure \ref{fig:means} shows the mean distributions for the four variables as a function of mean $Q$, compared for the two programs.

\begin{itemize}
\item Figure \ref{fig:means} a) shows the mean $\tau_z$ behaviour with increasing $Q$. This illustrates more clearly than the differential plots that the events become more collimated as $<Q>$ increases. The different points at the same $<Q>$ cover different $x$ bins; a clear $x$ dependence can be seen. Both programs show good agreement\footnote{There was an error in the my code which rendered the errors in $<\tau_z>$ meaningless; they have been omitted for clarity.} except for the $x$ dependence.
\item The mean $C$ parameter in figure \ref{fig:means} b) shows the same behaviour as thrust with $<Q>$. The agreement is very good although there is a clear problem in the two lowest $Q^2$ bins. Both programs show minimal $x$ dependence for this observable.
\item Figure \ref{fig:means} c) shows the mean jet-broadening. Above $<Q>$ of 10 GeV the agreement is very good. Below $<Q>$ of 10 GeV, DISENT shows a clear turnaround in the distribution which is not observed in DISASTER++.
\item The jet-mass mean distribution is shown in figure \ref{fig:means} d). There is in general good agreement with both programs showing little $x$ dependence. DISASTER++ has difficulties with divergences in two of the bins which require further investigation.
\end{itemize}

\section{Conclusions}

A comparison between two next-to-leading order programs, DISENT and DISASTER++, has been presented for four event shape variables in deep inelastic scattering. Differential and mean distributions have been considered. Both programs show good agreement over a wide kinematic range in the differential distributions, although jet-broadening shows some difference in the lower $Q^2$ region. The expected behaviour with increasing $Q^2$ is reproduced by both programs. The mean distributions show reasonable agreement except for jet-broadening where DISENT shows a clear turnaround in the distribution, which is not observed with DISASTER++.

\section{Acknowledgements}

I am grateful to A. Doyle, N. Brook and R. Waugh for guidance. Thanks also to T. Shah for help with the code, and to M. Seymour and D. Graudenz, the authors of the programs, for many useful discussions.

\end{document}